\documentstyle[aps,preprint]{revtex}
\topmargin = - 0.2 cm
\def\G{\Gamma}

\def\l{\lambda}

\def\pmb#1{\setbox0=\hbox{#1}%
  \kern-.025em\copy0\kern-\wd0
  \kern.05em\copy0\kern-\wd0
  \kern-.025em\raise.0433em\box0}

\def\bfxi{\pmb{$\xi$}}

\begin{document}

\begin{titlepage}
\begin{flushright}
MIT-CTP\#2362\\
IFUP-TH-54/94
\end{flushright}
\begin{center}
\Large\bf
Fermi-Walker coordinates in 2+1 dimensional gravity\footnote{This work is
supported in part by D.O.E. cooperative agreement
DE-FC02-94ER40818 and by M.U.R.S.T.}
\end{center}
\smallskip
\begin{center}
Pietro Menotti\\
{\small\it
Dipartimento di Fisica dell' Universit\'a, Pisa 56100, Italy and\\[-5.5pt]
I.N.F.N., Sezione di Pisa.}\\[-4pt]
{\small \tt e-mail: menotti@ibmth.difi.unipi.it}\\
{\small\rm  and}\\
Domenico Seminara\\
{\small\it
Center~for~Theoretical~Physics,~{Laboratory~for~Nuclear~Science~and~Department~of~Physics}\\[-4pt]
Massachusetts Institute of Technology, Cambridge, MA 02139 U.S.A.,}\\[-4pt]
\small\it
and I.N.F.N., Sezione di Pisa.\\ [-4pt]
{\small \tt e-mail: seminara@pierre.mit.edu}
\end{center}
\smallskip
\begin{center}
Talk presented at  Seventh Marcel Grossmann Meeting on General Relativity,\\
Stanford University, Stanford, California, U.S.A.\\
July 24-29, 1994
\end{center}

\end{titlepage}

\newpage
\baselineskip .5truecm
\begin{abstract}
\baselineskip .5truecm
It is shown that in 2+1 dimensions the Fermi-Walker
gauge allows the general solution of the problem of determining the
metric from the sources in terms of simple quadratures.
This technique is used to solve the problem of the occurrence of
closed time like curves (CTC's) in stationary solutions.
In fact the Fermi-Walker gauge, due to its physical nature, allows to
exploit the weak energy condition and in this connection it is
proved that, both for open and closed universes with axial
symmetry, the energy condition  imply the
total absence of closed time like curves.
The extension of this theorem to the general stationary problem, in
absence of axial symmetry is considered and at present the proof
of such generalization is
subject to some assumptions on the behavior of the determinant of the
dreibeins in this gauge.
\end{abstract}


\bigskip
\bigskip

\noindent
In the first order formalism the Fermi-Walker gauge is defined by
\begin{equation}
\label{form1}
\sum_i\xi^i\Gamma^a_{bi}=0~;~~~\sum_i\xi^i e^a_i=\sum_i\xi^i\delta^a_i.
\end{equation}
Regularity conditions at the space origin $\xi^i=0$ imply
\begin{equation}
\Gamma^a_{bi}({\bf 0},t)=0,~~~~~~~e^a_i({\bf 0},t)=\delta^a_i.
\end{equation}
Eqs.(\ref{form1}) are solved by the following formulas
\begin{equation}
\label{A1}
\G^a_{bi}(\xi)=\xi^j\int^1_0 R^a_{bji}(\lambda \bfxi, t)\lambda
d\lambda~;
{}~~\Gamma^a_{b0}({\xi})=\G^a_{b0}({\bf 0},t)+\xi^i\int^1_0 R^a_{bi0}
(\l {\bfxi},t) d\l,
\end{equation}
\begin{equation}
\label{A9}
e^a_0({\bf \xi})=\delta^a_0+\xi^i\int^1_0\Gamma^a_{i0}(\l{\bfxi},t)
d\l~;
{}~e^a_i(\xi)=\delta^a_i+\xi^j\int^1_0\G^a_{ji}(\l {\bfxi},t)\l d\l ,
\end{equation}
being $R^a_{bji}$ the curvature.

\noindent In $2+1$
dimensions a simplifying feature occurs because the Riemann
tensor, being a linear function of the Ricci tensor, can be written
directly in terms of the energy-momentum tensor
\begin{equation}
\label{einstein}
\varepsilon_{abc} R^{ab}= -{2\kappa} T_c.
\end{equation}
Thus also in the time
dependent case eqs.(\ref{A1},\ref{A9}) provide a
solution by quadrature of Einstein's
equations. Nevertheless one has to keep in mind that the solving
formulas are true only in the Fermi-Walker gauge, in which the energy
momentum tensor is not an arbitrary function of the coordinates but it is
subject to the covariant conservation law and symmetry property
that are summarized by the equations
\begin{equation}
{\cal D} T^a=0~~~~~{\rm and}~~~~~\varepsilon_{abc} T^b\wedge e^c =0.
\end{equation}
It will be useful as done in ref.\cite{MS1} to introduce the
cotangent vectors
$\displaystyle{T_{\mu}=\frac{\partial \xi^0}{\partial \xi^
\mu}}$,
$\displaystyle{P_{\mu}=\frac{\partial \rho}{\partial \xi^
\mu}}$ and
$\displaystyle{\Theta_{\mu}=\rho\frac{\partial \theta}{\partial \xi^
\mu}}$ where $\rho$ and  $\theta$ are the polar variables in the
$(\xi^1,\xi^2)$ plane. In addition  we notice that in $2+1$ dimensions
the most general form of a connection satisfying eq. (\ref{form1}) is
\begin{equation}
\label{connection}
\Gamma^{ab}_\mu(\xi)=\varepsilon^{abc}\varepsilon_{\mu\rho\nu}P^\rho
A^\nu_c(\xi).
\end{equation}
The covariant conservation equation is already satisfied while writing
$A^\nu_c$ in the form\cite{MS1}
\begin{equation}
\label{A}
A^\rho_c(\xi)= T_c \left [ \Theta^\rho\beta_1+T^\rho\frac{(\beta_2-1)}{\rho}
\right ]+\Theta_c \left [\Theta^\rho\alpha_1 +T^\rho\frac{\alpha_2}{\rho}
\right ]+P_c \left [\Theta^\rho \gamma_1+T^\rho\frac{\gamma_2}{\rho}
\right ],
\end{equation}
the symmetry constraint gives
\label{symmetryeq}
\begin{eqnarray}
\label{symmetry1}
&&A_1\alpha_2-A_2 \alpha_1+B_2
\beta_1-B_1\beta_2=0\\
\label{symmetry2}
&&A_2\gamma_1 -A_1
\gamma_2 +\frac{\partial B_1}{\partial\theta}-
{\partial B_2\over \partial t}=0\\
\label{symmetry3}
&&B_2\gamma_1-B_1
\gamma_2 +\frac{\partial A_1}{\partial\theta}-{\partial A_2\over
\partial t}=0,
\end{eqnarray}
where $A_1+1$, $B_1$, $A_2$, $B_2$ are the primitives of $\alpha_1$,
$\beta_1$, $ \alpha_2$, $\beta_1$.
The previous equations can be solved by means of simple quadratures
in the time depedent case in presence of rotational symmetry or in the
stationary case, in both cases  with complete control of the support
of the energy momentum tensor\cite{MS1}.
Moreover in the stationary case the projection
technique due to
Geroch\cite{geroch} allows to deduce from the completeness of such a
projection the completeness of the Fermi-Walker coordinate system.

\noindent
In the stationary case for example, the explicit solution by quadrature
given the $\alpha_1,\beta_1,\gamma_1$ is\cite{MS1}
\begin{equation}
\alpha_2={B^2_1\over B^2_1-A^2_1}
{\partial\over \partial\rho}\left ({N\over  B_1}\right )+
2\alpha_1 I
\end{equation}
\begin{equation}
\beta_2={A^2_1\over B^2_1-A^2_1}
{\partial\over \partial\rho}\left ({N\over A_1}\right )+
2\beta_1 I
\end{equation}
\begin{equation}
\gamma_2={B^2_1\over B^2_1-A^2_1}
{\partial\over\partial\theta}\left ({A_1\over B_1}\right )+
2\gamma_1 I,
\end{equation}
where
\begin{equation}
I=\int^\rho_0 d\rho^\prime {N (A_1\beta_1-B_1 \alpha_1)\over (B_1^2-
A_1^2)^2}~;~ N={1\over 2\gamma_1}{\partial\over \partial \theta} (A_1^2-B_1^2)
\end{equation}
while the support conditions for the energy momentum tensor are
\begin{equation}
\beta_1^2-\alpha_1^2-\gamma_1^2= {\rm const}.
\end{equation}
\begin{equation}
\alpha_1 B_1-\beta_1 A_1= {\rm const}.
\end{equation}
outside the sources. Such support equations are related to some Lorentz
and Poincar\'e holonomies\cite{MS1}.

\noindent
The techniques outlined above turns out to be a powerful tool  in
investigating the problem of the occurence of CTCs in 2+1
dimensions\cite{DJH,MS1,CFGO,tH1} .
In ref.\cite{DJH} the following ``Kerr'' solution in 2+1 dimensions was derived
\begin{equation}
\label{kerr}
ds^2=-(d t + 4 G J d\theta)^2+dr^2+(1-4 G M)^2 r^2 d \theta^2,
\end{equation}
which has the embarassing feature of having CTCs near the source.
The Fermi-Walker gauge due to its physical meaning allows to exploit
the weak energy conditions (WEC), to show that the functions
\begin{equation}
E^{(\pm)} (\rho)\equiv
(B_2\pm A_2)(\alpha_1\pm\beta_1)-
(\alpha_2\pm\beta_2)(B_1\pm A_1)
\end{equation}
are non increasing in $\rho$. From this result the following
theorem follows\cite{MS1}:\\

\noindent
For a stationary open universe with axial symmetry if the
matter sources satisfy the WEC and there are
no CTC at space infinity, then  there are no CTC at all. Thus the
``singular  source'' related to (\ref{kerr}) does not satisfy the WEC.\\

\noindent
With the same techniques the theorem on the absence of CTC's can also
be proved for all closed stationary universes with axial symmetry.
With regard to the extension to non axially-symmetric stationary
universes at present the proof goes through provided $\det(e)$ in the
Fermi-Walker gauge never vanishes.

\end{document}